\documentclass[superscriptaddress, twocolumn, amsmath,amssymb, aps, prl]{revtex4-1}

\usepackage{color}
\usepackage{amsmath}
\usepackage{graphicx}
\usepackage{dcolumn}
\usepackage{bm}

\newcommand{\GGG}{Gd$_3$Ga$_5$O$_{12}$}
\newcommand{\GAG}{Gd$_3$Al$_5$O$_{12}$}

\begin{document}

\title{Absence of magnetic ordering and field induced phase diagram in the Gadolinium Aluminium garnet}

\author{O. Florea}
\author{E. Lhotel}
\email{elsa.lhotel@neel.cnrs.fr}
\affiliation{Institut N\'eel, CNRS \& Universit\'e Grenoble Alpes, Grenoble, France }

\author{H. Jacobsen}
\affiliation{Niels Bohr Institute, University of Copenhagen, Copenhagen, Denmark}
\affiliation{European Spalliation Source, Lund, Sweden}
\affiliation{Department of Physics, Oxford University, Oxford, United Kingdom}

\author{C. S. Knee}
\altaffiliation{present address: ESAB AB, Lindholmsall\'en 9, SE-402 77 Gothenburg, Sweden.}
\affiliation{Department of Chemistry and Chemical Engineering, Chalmers University of Technology, Gothenburg, Sweden}

\author{P. P. Deen}
\email{pascale.deen@esss.se}
\affiliation{Niels Bohr Institute, University of Copenhagen, Copenhagen, Denmark}
\affiliation{European Spalliation Source, Lund, Sweden}

\begin{abstract}
The robustness of spin liquids with respect to small perturbations, and the way magnetic frustration can be lifted by slight changes in the balance between competing magnetic interactions remains a rich and open issue. We address this question through the study of the Gadolinium Aluminium garnet \GAG, a related compound to the extensively studied \GGG.  We report on its magnetic properties at very low temperature. We show that despite a freezing at about 300 mK, no magnetic transition is observed, suggesting the presence of a spin liquid state down to the lowest temperatures similarly to \GGG, in spite of a larger ratio between exchange and dipolar interactions. Finally, the phase diagram as a function of field and temperature is strongly reminiscent from the one reported in \GGG. This study reveals the robust nature of the spin liquid phase for Gd ions on the garnet lattice in stark contrast to Gd ions on the pyrochlore lattice for which a slight perturbation drives the compound into a range of magnetically ordered states.

\end{abstract}

\maketitle

In the last few decades, a great interest has been devoted to the study of frustrated systems. In particular, in geometrically frustrated magnetic systems, a competition between exchange energies is induced by the geometry of the lattice. It prevents conventional magnetic ordering, resulting in exotic correlations and ground states that can remain disordered down to zero temperature \cite{Lacroix11}.
Additional perturbations such as exchange interactions beyond the nearest neighbor exchange, dipolar interactions, quantum fluctuations or disorder can then play a crucial role and select a unique ground state in the system. The characterization of the robustness of the frustrated ground states with respect to these perturbations is thus an important question. 

\begin{figure}
\includegraphics[width=6.5cm]{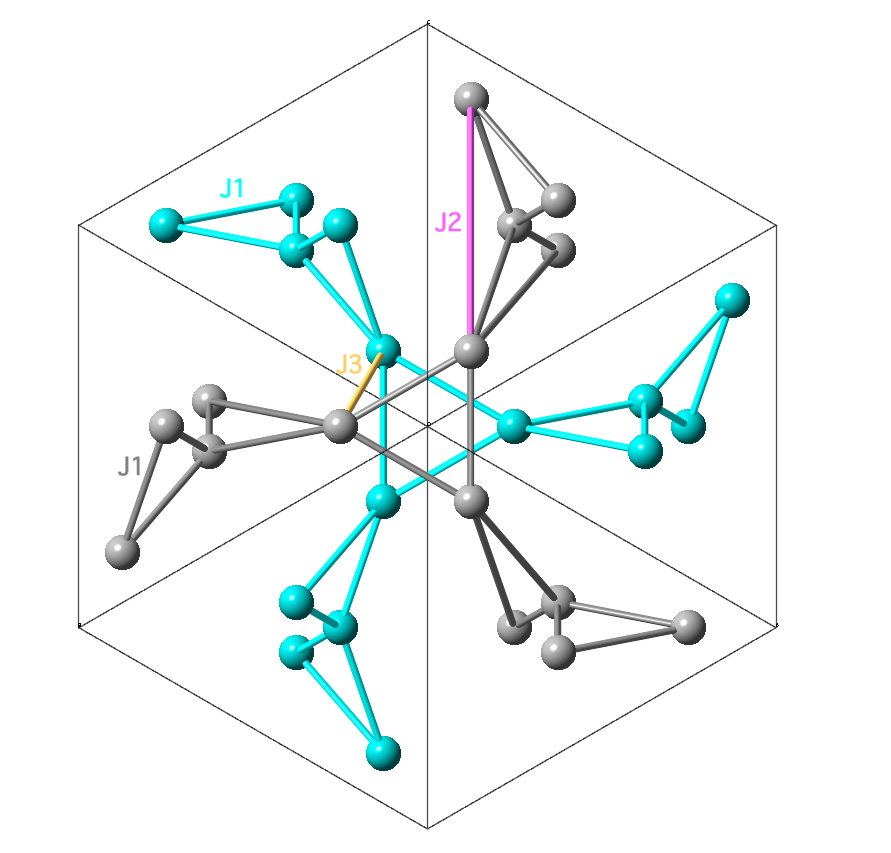}
\caption{\label{GAG_structure} 
Picture of the Gd$^{3+}$ ions in the GAG structure ($Ia{\bar 3}d$ space group): the spins lie on two interpenetrating three-dimensional hyperkagome lattices made of corner sharing triangles. The intralattice first ($J_1$) and second ($J_2$) neighbor exchange paths are represented as well as the interlattice ($J_3$) exchange path.  }
\end{figure}

Among the lattices that are prone to exhibit magnetic frustration, an interesting example is the hyperkagome structure, a three dimensional lattice of corner-sharing triangles. Few realizations of such structure have been discovered. The main examples are the garnets of formula $X_3A_2B_3$O$_{12}$ (with $X$ a magnetic element, and $A$, $B$ non magnetic elements), the iridate compound Na$_4$Ir$_3$O$_8$ \cite{Okamoto2007, ReferenceDally14, ReferenceShockley15} and more recently PbCuTe$_2$O$_6$ \cite{Mendels2016}. 
While the last two compounds are studied for their properties related to quantum effects, many studies on the frustration in magnetic garnets  have focused on the Gadolinium Gallium garnet GGG (formula \GGG), considered as an archetypal system to study the classical Heisenberg model with antiferromagnetic interactions on the hyperkagome lattice. The interesting physics in this compound is derived directly from the peculiar structure, Gd ions positioned on two interpenetrating hyperkagome lattices (see Figure \ref{GAG_structure}), and the effect of the long-range dipolar interactions on the ground state.    

Experimentally, GGG presents no evidence of conventional long-range ordering down to 25 mK \cite{ReferenceHov80}. 
Short-range correlations are observed down to the lowest temperature in neutron scattering experiments \cite{ReferencePetrenko98}. In addition to these correlations, longer correlations develop below 140 mK suggesting the existence of small ordered islands \cite{ReferencePetrenko98}, while a spin glass like behavior was reported below 200 mK in macroscopic magnetic measurements \cite{ReferenceSchiffer95}. Recently, the spin liquid state was shown to be associated with the development of correlated loops of 10 spins, whose resultant director moments are themselves long-range ordered \cite{ReferenceDeen15}. 
A rich phase diagram develops under applied magnetic fields, with the competition between antiferromagnetic, incommensurate and ferromagnetic field  induced phases \cite{ReferenceHov80, ReferenceSchiffer94, ReferencePetrenko08}, which converge to a single point at about 0.9~T and 350~mK \cite{ReferenceDeen-Florea15}. 

Some of these unconventional features were accounted for by including long range dipolar interactions, second and third nearest-neighbor interactions, raising the sensitivity of the ground state with respect to these parameters \cite{ReferenceKinney79, ReferencePetrenko99, ReferenceYavorskii06}.
Indeed, while GGG does not order down to the lowest temperatures, similar magnetically isotropic garnets, Mn$_3$Al$_2$Ge$_3$O$_{12}$ and Mn$_3$Al$_2$Si$_3$O$_{12}$ do order \cite{ReferencePrandl73}. In these systems, the ordering was attributed to the strong inter sublattice interactions ($J_3$ in Figure~\ref{GAG_structure}) which lifts the degeneracy of the ground state \cite{ReferenceValyanskaya76}. In GGG, $J_3$ is much smaller, thus emphasizing the role of the dipolar interaction. 

To probe more specifically the sensitivity of the GGG phase diagram with respect to variations in the magnetic couplings, we focus here on another Gadolinium garnet \GAG\ (GAG), which has the same hyperkagome structure as GGG but a smaller lattice parameter due to the smaller ionic radii of Al$^{3+}$ compared with Ga$^{3+}$ ions \cite{ReferenceQuilliam13}.  It allows to increase the antiferromagnetic exchange interactions by keeping the same magnetic ion and structural properties \cite{ReferenceHamilton14}, while the dipolar interactions remain almost constant, making GAG a perfect candidate to address this question.
Contrary to GGG, a small peak has been reported in specific heat measurements of GAG at 175 mK, which was attributed to long-range ordering \cite{ReferenceQuilliam13}. 

In this Letter, we present the first magnetic study of GAG at very low temperature and we map its field-temperature phase diagram. We do not observe signatures of conventional ordering, and we show that, contrary to expectation, GAG's ground state as well as its phase diagram are very similar to the ones of GGG, the energy scales being renormalized by the nearest neighbor interaction. Our study thus points out the robustness of the spin liquid state and of the field induced phases in the classical hyperkagome Gd garnet family. 

A polycrystalline sample was prepared using a sol-gel synthesis route as outlined in Ref. \onlinecite{Garskaite2007}, using high purity Al(NO$_3$)$_{3}.9$H$_2$O and isotopically enriched $^{160}$Gd$_2$O$_3$. The enrichment level of the gadolinium oxide was 98.4~\%. The sample was subjected to a final heat treatment at $1000~^{\circ}$C and was judged to be phase pure from long scan X-ray diffraction (XRD) data.  A cell parameter, $a = 12.1217(1)$ \AA\ was obtained from Rietveld analysis of the XRD data.

We performed AC susceptibility and magnetization measurements on this sample down to 100 mK. In the ``high" temperature range, $2-300$ K, a commercial Quantum Design SQUID magnetometer was used. Below 4 K,  we used two SQUID magnetometers developed at the Institut N\'eel and equipped with $^3$He-$^4$He dilution refrigerators \cite{ReferencePaulsen01}. Magnetic fields up to 8 T were applied.  In the low temperature measurements, the sample was mixed with apiezon N grease to ensure thermalization, and to prevent the sample from reorienting in the presence of a magnetic field. The temperature sweeping rate was 2 mK/min or 5 mK/min below 400 mK, which is slow enough to measure reversible curves, and thus to be at thermal equilibrium.

 \begin{figure}
	\includegraphics[width=8cm]{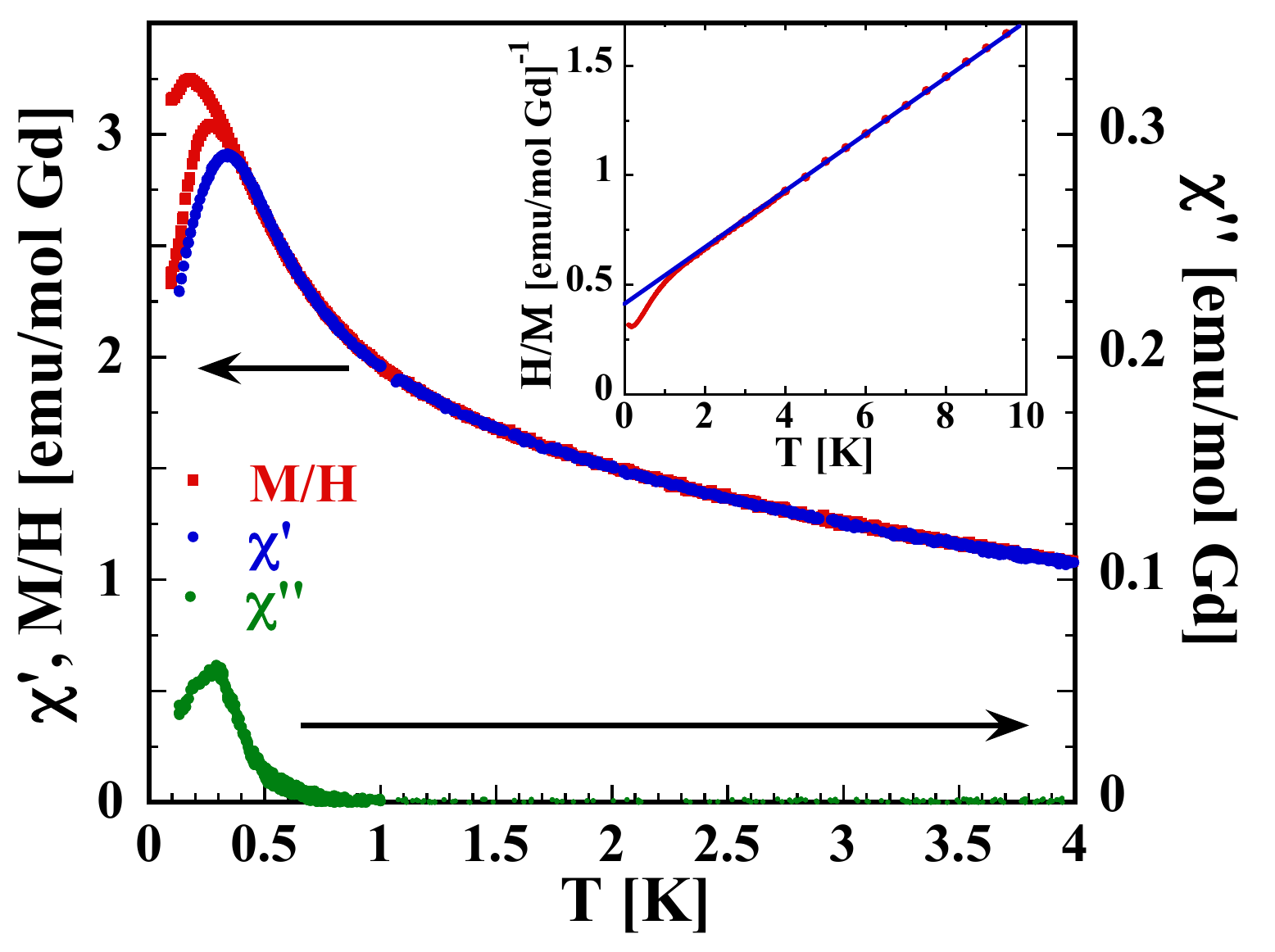}
	\caption{\label{GAG_MT_50G} Magnetization $M/H$ (red points) and AC susceptibility vs $T$. $\chi'$ (blue points) and $\chi''$ (green points) are the in phase and out of phase parts of the AC susceptibility. For the magnetization, the applied DC field is 5 mT. The AC susceptibility was measured in zero DC field with $\mu_0H_{\rm AC}=0.55$~mT and $f=0.57$ Hz. Inset: $H/M \approx 1/\chi$ vs $T$ for temperatures between 100 mK and 10 K. The blue line is a fit to a Curie-Weiss law between 20 and 300 K: $1/\chi = 0.499 + 0.128~T$. }
\end{figure}

In the high temperature regime, above 2 K, we recover previous results by Hamilton et al. \cite{ReferenceHamilton14}: the susceptibility follows a Curie-Weiss law down to 20 K. Below this temperature, the susceptibility slightly deviates from the Curie-Weiss behavior. The fit of the inverse susceptibility between 20 and 300 K gives a Curie constant of  7.8 emu.K/mol which is consistent with a spin equal to $S=7/2$ as expected for the Gadolinium ion. The Curie-Weiss temperature $\theta_{\rm CW} = -3.9 \pm 0.3$ K, is larger than in GGG ($\theta_{\rm CW} \sim -2.25$~K \cite{ReferenceKinney79}). Assuming $\theta_{\rm CW}=\dfrac{J_1 n S(S+1)}{3k_B}$, with $J_1$ the nearest neighbor interaction and $n=4$ the coordination number, gives $J_1=186$~mK (consistent with Ref. \onlinecite{ReferenceHamilton14}), compared to $J_1=107$ mK in GGG \footnote{Ref. \onlinecite {ReferenceHamilton14} obtains a slightly larger Curie-Weiss temperature in GGG, giving $J_1=124$ mK. Note that theoretically, $\theta_{\rm CW}$ also contains the $J_2$ and $J_3$ exchange parameters. $\theta_{\rm CW}$ depends slightly on the temperature range of the fit and on the element diamagnetic contributions, which explains the uncertainty. Demagnetization corrections (which cannot be performed accurately for the present sample) would tend to decrease $\theta_{\rm CW}$, but within the given error bar}.
Considering the dipolar interaction $D$ limited to the nearest neighbors separated by a distance $r_{\rm nn}$, $D=\dfrac{\mu_0 \mu_{\rm eff}^2}{r_{\rm nn}^3}$ is equal to 48~mK in GAG, and 45 mK in GGG, so that the ratio $J_1/D$ is about 1.6 times larger in GAG than in GGG. 

Figure \ref{GAG_MT_50G} shows the magnetization measured in a zero field cooled - field cooled (ZFC-FC) procedure in a magnetic field of 5 mT: first, the sample is cooled in zero field and the magnetic field is applied at low temperature. The magnetization is measured while heating up to 1.5 K (ZFC curve) and then while cooling down to 100 mK (FC curve). The magnetization is measured heating up again to ensure the reversibility of the FC curve.
The inverse of the magnetization (see inset of Figure~\ref{GAG_MT_50G}) varies almost linearly as a function of temperature down to 1 K. 
Below this temperature, it clearly deviates from the paramagnetic behavior, and the magnetization increases faster than the Curie-Weiss law, suggesting the onset of ferromagnetic correlations. 
Below 270 mK, an irreversibility is observed between the FC and ZFC curves. 
This irreversibility is associated with a broad peak in both the real and imaginary parts of the AC susceptibility at about 300 mK. These features indicate that, similarly to GGG, a freezing occurs at low temperature. At the freezing temperature, which is ten times below the Curie-Weiss temperature, the system is already correlated. This effect thus may be related to some rigidity of magnetic clusters. The frequency dependence of the AC susceptibility however does not follow a dynamic scaling law, as would be expected for a canonical spin glass behavior \cite{SouletiePRB_1985}. 

A peak was reported in the specific heat in Ref. \onlinecite{ReferenceQuilliam13}, suggesting long-range ordering. Our magnetization data do not show evidence for this ordering: although a small peak is observed in the FC magnetization at about 175~mK, no anomaly is observed both in the ZFC magnetization and the AC susceptibility at this temperature as should be expected in the presence of a magnetic transition towards a long-range ordered state. In addition, preliminary neutron scattering measurements at about 60 mK do not provide evidence for conventional ordering. Very low temperature specific heat measurements should be performed to check whether the peak in the specific heat reported by Quilliam et al. \cite{ReferenceQuilliam13} is present in our sample. 

From our measurements in low magnetic fields, we can conclude that GAG presents a disordered correlated state down to 100 mK. A freezing is observed below 300~mK, similarly to GGG, but at a slightly larger temperature. Their ground states thus look similar in spite of the stronger $J_1$ interaction in GAG. We now address the way the field induced properties are modified in GAG with respect to GGG. 

\begin{figure}
\includegraphics[width=8cm]{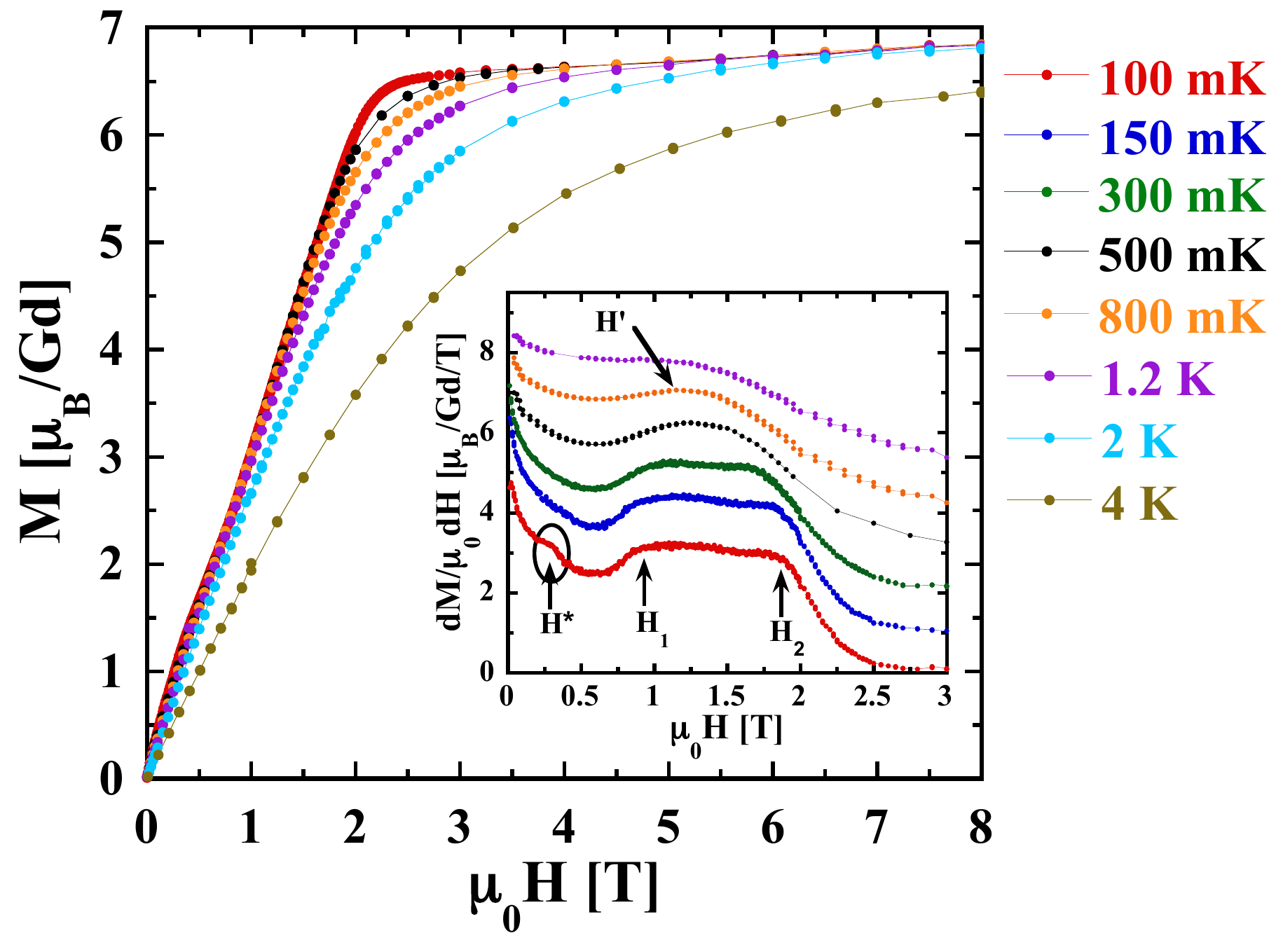}
	\caption{\label{GAG_dMdH} $M$ vs $H$ for temperatures between 100 mK and 4 K. Inset: $dM/dH$ vs $H$. The curves for different temperatures are shifted by steps of $1~\mu_{\rm B}/$T for clarity.}
\end{figure}

Isothermal magnetization measurements as a function of magnetic field performed for several temperatures between 100 mK and 4 K are displayed in Figure \ref{GAG_dMdH}. At 100~mK, the magnetization shows a saturation above about 3~T. Nevertheless, the saturated value is slightly below the expected saturated value $M_{\rm sat}=7~\mu_{\rm B}/{\rm Gd}$, while a small but finite slope remains up to 8 T  (Note that our absolute values at 4 K are in agreement with measurements of Ref. \cite{ReferenceHamilton14}): about 3\% of the magnetization is still missing at 8 T at 100 mK. This indicates that a part of the magnetic moment does not align with the applied field.
This effect was already pointed out in GGG, where it was shown to be associated with a canting of the moments due to the long-range dipolar interactions  \cite{ReferencedAmbrumenil15}.

At lower field, and for temperatures below 1.2 K, an anomalous curvature is observed in the magnetization curves, as emphasized in the inset of Figure \ref{GAG_dMdH} which shows the derivative of the magnetization with respect to the field as a function of field.
Indeed, as the sample is cooled down below 1.5~K, a broad maximum develops in $dM/dH$ around $\mu_0 H^{\prime} =1.5$~T. By further cooling down, this broad bump transforms into a plateau, delimited by the fields labeled $H_1$ and $H_2$. By comparing these results with the $dM/dH$ curves in GGG \cite{ReferenceSchiffer94, ReferenceHov80, ReferenceDeen-Florea15}, we can attribute this plateau to the presence of a field induced antiferromagnetic phase. The sharp features at $H_1$ and $H_2$ present in GGG single crystals are not observed, because our measurements are performed on a powder sample, so that the transition is broadened due to averaging over all directions and the presence of anisotropy \cite{ReferenceFloreaPhd, ReferenceHov80}.  
At the lowest temperature, an anomaly can be seen at $\mu_0 H^*= 0.25$ T, which was not present in GGG. In these curves, we do not observe any evidence of the peak observed between $H_1$ and $H_2$ in GGG for $H$ applied along the $[110]$ direction \cite{ReferenceDeen-Florea15}. Again, this can be due to powder averaging and thus precludes any further conclusions.

\begin{figure*}
	\includegraphics[width=\textwidth]{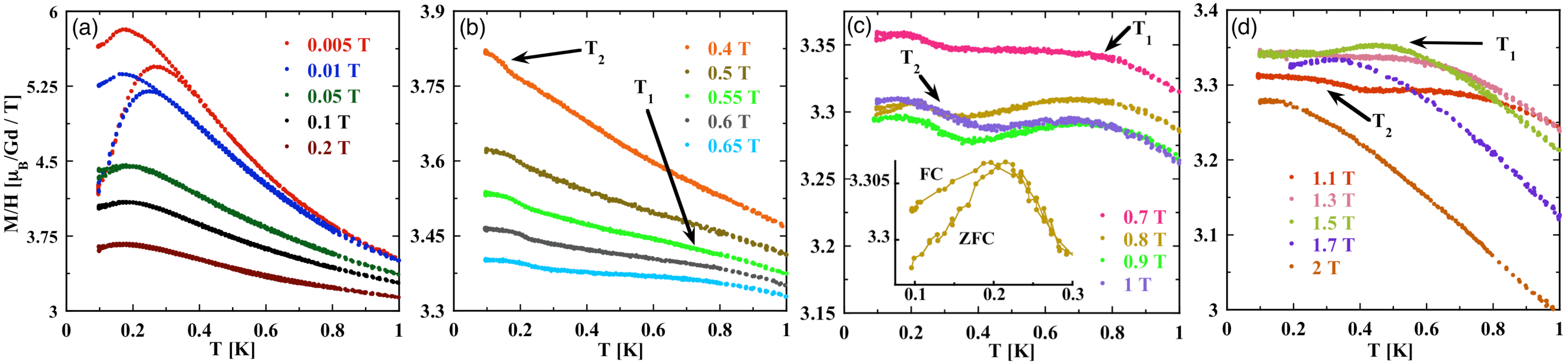}
	\caption{\label{GAG_MT} $M/H$ vs $T$ for several applied magnetic fields between 5 mT and 2 T measured in a ZFC-FC procedure. (a) from 5 mT to 0.2 T. (b) from 0.4 to 0.65 T. (c) from 0.7 to 1 T. Note the suppression of the ZFC-FC splitting as the applied magnetic field is increased and the re-entrant effect at $0.7-0.8$ T, highlighted in the inset. (d) from 1.1 to 2 T. }
\end{figure*}

To probe more accurately this field induced phase diagram, magnetization measurements as a function of temperature at constant applied fields were performed. All measurements were made using the ZFC-FC protocol, as described above.
Figure \ref{GAG_MT} displays the rich and complex behavior resultant from the measurements. 
Starting from the low field magnetization shown in Figure \ref{GAG_MT_50G}, upon increasing the field, the first consequence is a closure of the ZFC-FC irreversibility. It shows that this effect is associated with small energy scales in the system. At 0.1~T, the hysteresis has almost disappeared. A broad peak is still present and moves towards lower temperatures (Figure~\ref{GAG_MT}(a)). 
When further increasing the field up to 0.5~T, the peak disappears, and  the magnetization decreases monotonically as the temperature is increased (Figure \ref{GAG_MT}(b)). Nevertheless, at about 0.18 K, a change of slope occurs, which is marked by $T_2$ on the figure. Concomitantly, at larger temperature, a broad maximum starts to develop at $T_1$, and is further evidenced at larger fields up to 1.7~T. Above 0.7~T, the low temperature anomaly at $T_2$ is followed by a peak which is associated with a reentrant ZFC-FC irreversibility in the $0.7-0.8$ T field range (Figure \ref{GAG_MT}(c)). While a splitting of the ZFC-FC curves is quite common in low field in the presence of energy barriers, this reentrant effect is highly uncommon in higher fields. A similar feature was also observed in GGG \cite{ReferenceDeen-Florea15}, but its origin was not understood.
Above 0.9~T, both features at $T_1$ and $T_2$ shift to lower temperature when increasing the field, so that the $T_1$ anomaly disappears above 1.1 T, and the $T_2$ peak disappears at 2~T (Figure \ref{GAG_MT}(d)). 

\begin{figure}
	\includegraphics[width=8.5cm]{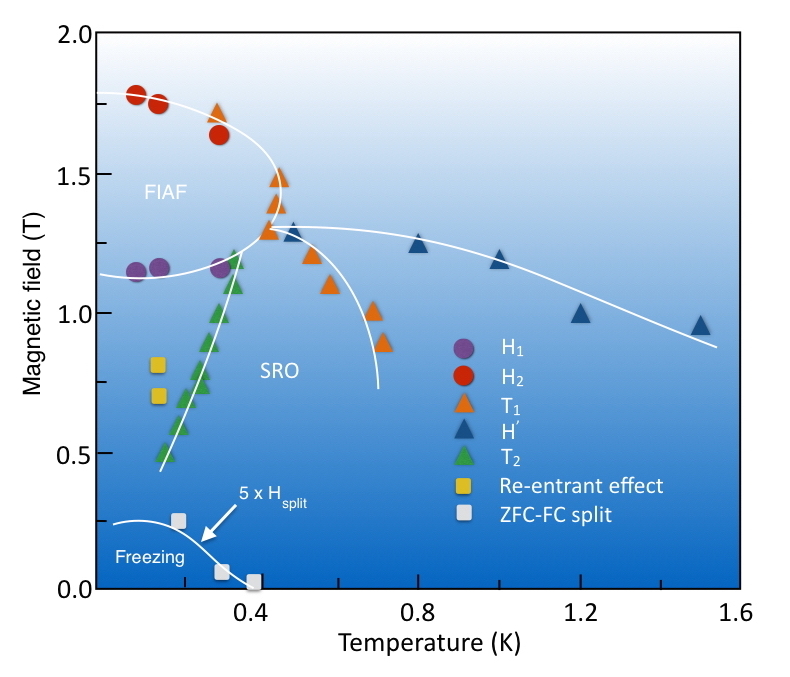}
	\caption{\label{GAG_phase_diagram} Magnetic field - Temperature phase diagramÊof GAG refined from magnetization measurements. The $H_{\rm split}$ values are scaled by a factor of 5 for visibility. }
\end{figure}

A global overview of the GAG $H-T$ phase diagram is presented in Figure \ref{GAG_phase_diagram}.
While our magnetization measurements alone prevent an accurate determination of the nature of the phases we have measured, the similarity with the observations in GGG, allows the development of a clear picture of the magnetic phases present in the phase diagram. 
Below 1.5 K, a short-range ordered (SRO) phase develops which seems to display several regimes, highlighted by the existence of a broad peak at $T_1$ and a change of slope in the temperature dependence of the magnetization at $T_2$.
Under applied magnetic field a field induced antiferromagnetic (FIAF) phase  occurs between 1.1 and 1.7 T, for temperatures below 450 mK. 
The short-range order line and the field induced phase seem to converge at 450 mK and 1.3 T. In comparison, the analogous point in GGG converges at 350 mK and 0.9~T. 
In this respect, the GAG phase diagram seems to have been shifted in temperature and field by a factor comparable with the increase of the nearest neighbor exchange $J_1$. This suggests that the boundaries of the field induced phase diagram in the Gd garnet family are governed by $J_1$. Other couplings are however expected to influence the exact structure and dynamics in each phase \cite{ReferencedAmbrumenil15}. 
 
In GAG, the re-entrance of the ZFC-FC hysteresis occurs in the $0.7-0.8$~T range, i.e. outside of the field induced phase. This contrasts with GGG, where this reentrance was found deeply in this FIAF phase when the field is applied along the $[110]$ direction. In that case, it is associated with a peak in the $dM/dH$ vs $H$ curves \cite{ReferenceDeen-Florea15} and an anomaly in sound velocity measurements \cite{Rousseau17}
This suggests that the mechanism of this re-entrant behavior is not directly related to $J_1$, neither with the ``converging" point as was initially argued in GGG. A careful neutron scattering study around this field in GAG as well as in GGG might help capturing the origin of this phenomenon. 

We have thus shown that GAG presents a zero field behavior similar to the GGG spin liquid, but which manifests at larger temperature due to the larger nearest neighbor coupling. In applied magnetic field, we recover the signatures of short-range ordering above 450 mK, and of field induced ordering at lower temperature. Given the much larger value of $J_1/D$ in GAG compared to GGG, our study thus points out that the phase diagram of GGG is robust with respect to strong changes in the ratio between nearest neighbor and dipolar interactions. 
This robustness is at variance with what is observed in other Gd based frustrated materials, e.g. the pyrochlore oxides: the Gd$_2$Ti$_2$O$_7$ and Gd$_2$Sn$_2$O$_7$ compounds, although they stabilize a long-range ordered ground state, exhibit totally different magnetic structures, with different propagation vectors \cite{Wills_JPCM06, ChampionPRB_2001}. In these cases, the difference was attributed to the role of second and third neighbor interactions. In the Gadolinium garnet family, it appears that contrary to expectations, the spin liquid state is a robust ground state, weakly sensitive to the perturbations of the coupling parameters. Further neutron scattering measurements will provide a deeper understanding of the correlated phase in GAG and will allow to better refine the whole characteristics of the phase diagram we have described.     

\acknowledgements
We thank C. Paulsen for the use of his low temperature magnetometers. O.F. acknowledges funding from the Laboratoire dÕexcellence LANEF in Grenoble (ANR-10-LABX-51-01).

\bibliographystyle{apsrev4-1}
\bibliography{Bibliography_GAG}

\end{document}